\documentclass[12pt]{iopart}
 \usepackage{amsfonts,amssymb,mathtools}
\usepackage{graphicx}
\usepackage{subcaption}
\usepackage{epstopdf}

\DeclareUnicodeCharacter{2212}{-}

\begin{document}
 \title{Investigating the interaction of a Cosmic String with an Accreting Black Hole}
 \author{I Swamy, D Singh}
 \address{Department of Physics, Dr. Vishwanath Karad MIT-World Peace University, Kothrud, Pune, India, 411038}

 \ead{deobrat.singh@mitwpu.edu.in}

 \begin{abstract}
 Rotating black holes when attached to a cosmic string have their rotational energy extracted leading to a change in its spin and mass. The spin of a black hole can be measured using various methods for an accreting black hole in an X-ray binary system. Accretion disks around black holes have an innermost stable circular orbit (ISCO) whose location is directly dependent on spin and mass of the black hole. The orbit's location changes as the black hole's spin changes and hence can be a method to detect the presence of cosmic strings. This study investigates this change and suggests the ejection of accretion material as black hole spin approaches maximum for prograde motion and material falling into the black hole for retrograde motion, regardless of the presence of cosmic string. However, in the presence of cosmic string, the spin-up process due to accretion is found out to be slower, even with high accretion rates and is detectable. There is a transition phase that occurs as the black hole approaches maximum spin, where even small changes in spin result in significant changes in the ISCO's position. Accreting black holes attached to a large string never reach this transition phase and this absence serves as potential evidence for the existence of a cosmic string. 
 \end{abstract}
 \submitto{\PS}
 \maketitle

\section{Introduction}
Symmetry breaking phase and topological defects in the early universe lead to the formation of one dimensional string-like energy densities, known as cosmic strings \cite{Kibble76}\cite{Vilenkin85}. These are hypothetical structures that arise as solutions of field theories and string theory \cite{Polchinski04}. Gravitational lensing \cite{Sazhin03} and gravitational waves \cite{Vachaspati85} are currently the prominent methods to detect the existence of cosmic strings, with gravitational wave data from LIGO setting an upper limit of $10^{-11}$ on the cosmic string tension \cite{Abbot21}. 
The interaction of cosmic strings with black holes has been a major research topic and provides for interesting phenomena (for more details refer \cite{Larsen96}). Primordial black holes formed in the early universe form with strings attached to it, resulting in black-hole-string networks \cite{Vilenkin18}. Other studies suggest that string interaction with the gravitational field of the black hole results in its chaotic capture, proposing chaotic scattering and unstable periodic orbits around the black hole as possible effects of this chaotic capture \cite{Larsen94}\cite{Larsen98}\cite{Dubath07}\cite{Larsen93}. The case of charged circular string is also studied which suggests the existence of stable stationary strings due to a critical charge current \cite{Frolov99}.  
Rotating black holes attached to a cosmic string has also been studied, with \cite{Xing2021} showing the production of smaller string loops bound to the black hole. The attached cosmic string also leads to rotational energy of the black hole being extracted by the string\cite{Kinoshita16}, eventually leading to its spin-down\cite{ahmed24}. The spin of black holes can be detected through gravitational waves, but at present, this is only possible for binary black holes \cite{Abbott16}\cite{Roulet19}. 
\\

This spin-down process becomes observable for an accreting black hole, predominantly found in X-Ray binary systems (XRB).  These are systems consisting of a companion star (main sequence star, white dwarf, red giant or supergiant) orbiting a compact star (neutron star or a black hole) \cite{Tauris06}. XRBs are classified, based on the mass of the companion star, into Low Mass X-ray binaries (LMXB) and High Mass X-ray binaries (HMXB). Black Hole XRBs have black holes accreting material from its companion star, forming a disk-like structure. This system has been studied in a recent work by the authors suggesting that XRBs provide for a better detection method by observing the impact the energy extraction by the string has on the orbital period of the companion star \cite{Swamy25}.
\\

The spin of accreting black holes can be estimated by observing its jet, formed due to the presence of a magnetic field around the black hole \cite{Blandford77} with observations backing the relation between jet power and spin \cite{Narayan12}. However, for accreting black hole without an observable jet, quasi periodic oscillations (see \cite{Ingram19} for a review) serve as a possible method to detect spin. 
\\

The orbits of the materials in the disk depends on the spin of the black hole and hence provide for an excellent method to estimate the black hole spin. The innermost stable circular orbit (ISCO) is the smallest orbit where a test particle can orbit stably without spiralling into the black hole \cite{Bardeen72}. Consequently, the change in a black hole's spin due to a cosmic string can be observed by the location of the ISCO(see \cite{Reynolds21} for a review). This location can be determined by assesing the shift in the iron lines of this orbit's spectrum \cite{REYNOLDS03} or by employing a multitemperature blackbody model \cite{Li09}\cite{Yuan04}\cite{Shcherbakov12}\cite{Gou11}.\\

The motivation behind this study is to understand the effect the cosmic string has on the structure of the accretion disk due to the rotational energy extraction by the string and provide for a potential detection method for cosmic strings. We will analyse the impact of a cosmic string on the accretion disk orbits of a black hole in an LMXB. To investigate this impact, LMXBs have been preferred over HMXBs for two major reasons: firstly, the low mass of the star makes it possible to neglect its gravitational interaction with the string and secondly, its lifetime being greater than an HMXB gives the black hole enough time to accrete and have an observable spin change. Section 2 explores the change in the location of ISCO for changing mass and spin of a black hole and further assess the effect a cosmic string has on the location of this orbit. In Section 3, a numerical analysis is conducted on the equations shown in Section 2 and its results are discussed in Section 4. We propose here that for cosmic strings with both small and large string tensions, the effect on the black hole spin and accretion disk is observable, with the effect being more observable for larger string tensions.

\section{Radius of Innermost Stable Circular Orbit}
\begin{figure}[h!]  
   \centering
  \includegraphics[width=0.8\linewidth, height = 7 cm]{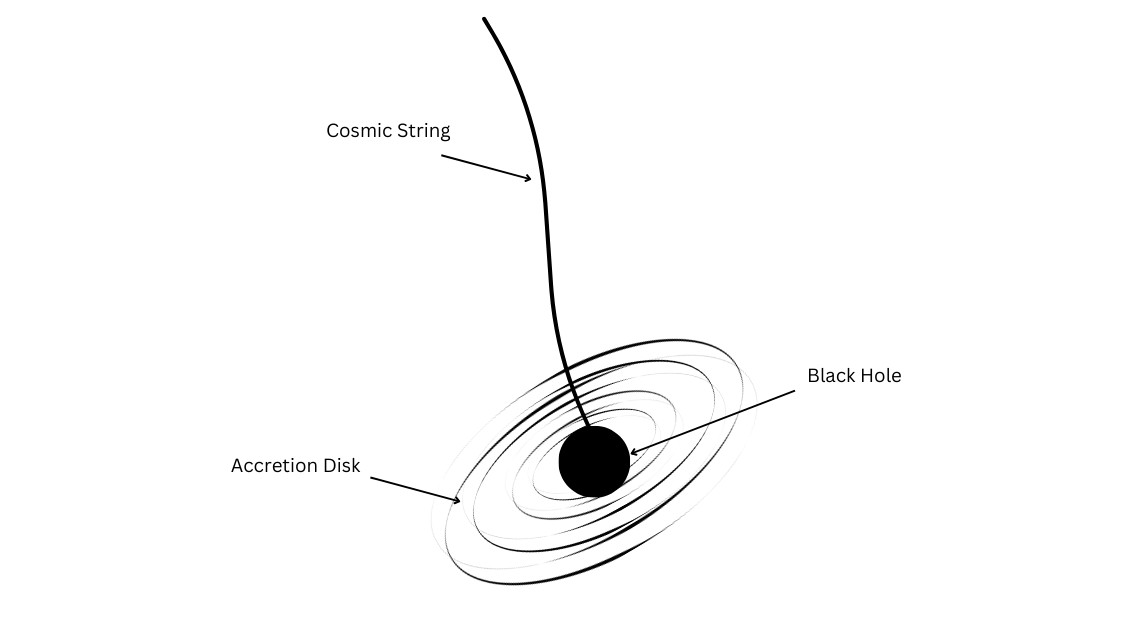}
  \caption{Schematic illustration of the Black Hole - Cosmic String system considered.}
  \label{fig}
\end{figure}

We consider a Nambu-Goto string attached to an accreting black hole in a LMXB with one end attached and the other end extending out as depicted in Fig \ref{fig} (not to scale). The invariant length of the string is very large compared to the black hole radius, and hence effects such as oscillations are not being considered here. The gravitational interaction of the string and the star has been considered negligible due to the star's low mass. 
The dynamics of a rotating black hole solely depends on its mass and angular momentum and hence a spin parameter known as dimensionless Kerr parameter can be defined for it \cite{Kerr63}\cite{Misner73}. For a Black Hole of mass M, the dimensionless Kerr parameter $\alpha = J/M^2$ ranges from -1 to 1. Thus, the radius of ISCO is given as (in geometric units where G = c = 1) \cite{Bardeen72},
\begin{equation}
R_{ISCO} = M\{3 + Z_2 \mp [(3 - Z_1)(3 + Z_1 + 2Z_2)]^{1/2}\}
\label{bardeen}
\end{equation} 
where
\begin{equation}
  Z_1 \equiv 1 + (1 - {\alpha}^2)^{1/3}[(1 + \alpha)^{1/3} + (1 - \alpha)^{1/3}]  
\end{equation}
and
\begin{equation}
Z_2 \equiv (3{\alpha}^2 + {Z_1}^2)^{1/2}
\end{equation}
 We take
\begin{equation}
Z_0 \equiv 3 + Z_2 \mp [(3 - Z_1)(3 + Z_1 + 2Z_2)]^{1/2}
\label{z0}
\end{equation}
such that
\begin{equation}
    R_{ISCO} = MZ_0
\end{equation}
\\ We now attempt to find the change in $R_{ISCO}$ due to the changing spin of the Black Hole and hence for the first time establish a relation between $R_{ISCO}$ and the string tension $\mu$. In this regard, differentiation of (\ref{bardeen}) with respect to time leads to
\begin{equation}
    \dot{R}_{ISCO} = [\dot{Z}_0M + Z_0\dot{M}]
    \label{general}
\end{equation}
which gives the generalised change in $R_{ISCO}$ for a black hole with varying mass and spin. Obviously
\begin{equation}
    \dot{Z_0} = \dot{Z_2} \mp \frac{3\dot{Z_2} - Z_1\dot{Z_1} - Z_1\dot{Z_2}}{[(3 - Z_1)(3 + Z_1 + 2Z_2)]^{1/2}},
    \label{dotz0}
\end{equation}
\begin{equation}
    \begin{aligned}
        \dot{Z_1} = &\frac{-2\alpha\dot{\alpha}}{3(1-{\alpha}^2)^{2/3}}[(1 + \alpha)^{1/3} + (1 - \alpha)^{1/3}]\\ & + \frac{1}{3}\left[\left({\frac{1+\alpha}{1-\alpha}}\right)^{1/3} - \left({\frac{1+\alpha}{1-\alpha}}\right)^{1/3}\right],
    \label{dotz1}
    \end{aligned}    
\end{equation}
and
\begin{equation}
    \dot{Z_2} = \frac{3\alpha\dot{\alpha} + Z_1\dot{Z_1}}{(3{\alpha}^2 + {Z_1}^2)^{1/2}}.
    \label{dotz2}
\end{equation}
It is obvious from (\ref{general}) that the evolution with time not only depends on the spin parameter but also the mass of the black hole. This mass change is due to the accretion material adding to the black hole mass, while the mass loss could be due to jets and winds. 
Further exploring (\ref{general}) for an accreting black hole attached to a cosmic string, we must consider the mass loss through the cosmic string extraction process. Thus the change in mass can be explicitly written as 
\begin{equation}
\dot{M} = \dot{M}_{acc} - \dot{M}_{str}
\end{equation}
which implies
\begin{equation}
    \dot{R}_{ISCO} =  [\dot{Z}_0M + Z_0(\dot{M}_{acc} - \dot{M}_{str})]
    \label{specific}
\end{equation}
where $\dot{M}_{acc}$ is the increase in black hole mass due to accretion and $\dot{M}_{str}$ is the mass loss due to cosmic string. 
\\The acretion mass rate with radiative correction has been stated as \cite{Novikov73}\cite{talbot21},
\begin{equation}
 \dot{M}_{acc} = (1 - \epsilon_r)\dot{M}_{acc,0}   
 \label{accretion mass}
\end{equation} 
where
\begin{equation}
     \epsilon_r = 1 - \sqrt{(1 - \frac{2}{3Z_0})}
\end{equation} is the spin-dependent radiative efficiency and $\dot{M}_{acc,0}$ is the rest-mass flux on to the black hole.
As previously discussed, the string extracts the black hole's rotational energy, leading to a mass loss \cite{Kinoshita16}. The mass loss due to cosmic string can be approximated as \cite{ahmed24}
\begin{equation}
 \dot{M}_{str} \approx 10^4 \mu(M_{\odot}/s)   
 \label{string mass loss}
\end{equation}
where $\mu$ is the dimensionless string tension.
From (\ref{accretion mass}), (\ref{string mass loss}) and (\ref{specific}),
\begin{equation}
   \dot{R}_{ISCO} = [\dot{Z}_0M + Z_0\{(1 - \epsilon_r)\dot{M}_{acc,0}  - 10^4\mu\}] 
   \label{final}
\end{equation} 
where all the mass rates are taken in units $M_{\odot}/s$.
It is evident from expression (\ref{final}) that the change in ISCO is directly dependent on the string tension.  

\section{Numerical Analysis}
In this section, we primarily analyse the behaviour of $ \dot{R}_{ISCO} $ with changing spin ($-1<\alpha<1$) for different string tensions. We consider a black hole of 10$M_\odot$ accreting at a rate $\dot{M}_{acc} = 10^{-8}M_\odot/yr$.  We restate here that values of $\dot{R_{ISCO}}$ and  $R_{ISCO}$ are in geometric units (G = c = 1).

\subsection{Small String tension}
In the case of small string tension we take the value of $\mu$ of the order of $10^{-40}$ and inserting it into (\ref{string mass loss}) gives $\dot{M}_{str} = 10^{-29} M_{\odot}/yr$. The change in spin is taken to be $\dot{\alpha} = 10^{-7}/yr$. This value of $\dot{\alpha}$ has been selected on the basis that the timescale of accretion in LMXB is around $10^7 - 10^9 yr$ \cite{Tauris06} and hence the spin-up process is taken to be on a timescale of the same order.  Using this value and the previously defined values, in (\ref{dotz0}),(\ref{dotz1}),(\ref{dotz2}) and (\ref{specific}), the following plots have been generated in python 3.11.5.

\begin{figure}[h!]
     \begin{subfigure}{0.5\textwidth}
        \includegraphics[width=1\linewidth, height=6.5cm]{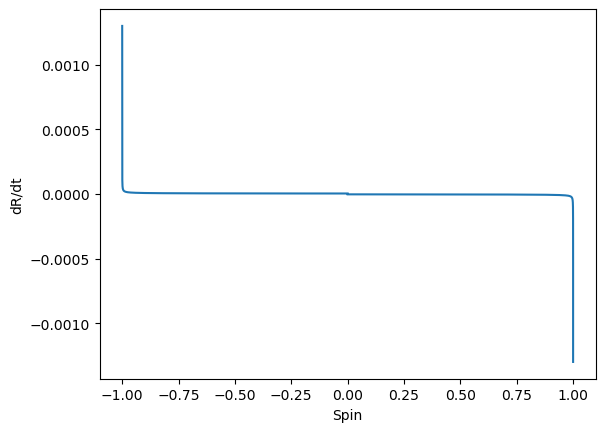} 
        \caption{$-1<\alpha<1$}
        \label{fig1a}
     \end{subfigure}
    \begin{subfigure}{0.5\textwidth}
        \includegraphics[width=1.1\linewidth, height=6.5cm]{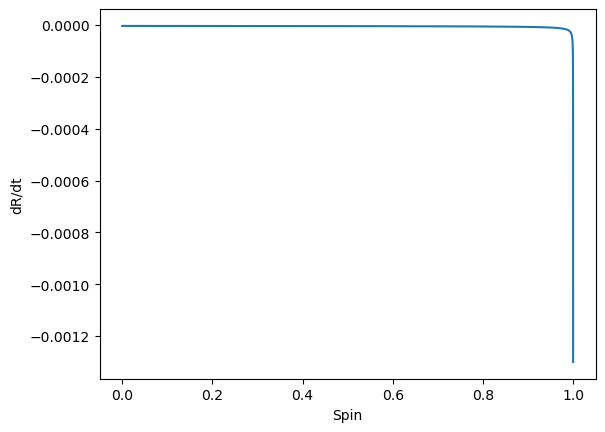}
        \caption{$0<\alpha<1$}
        \label{fig1b}
    \end{subfigure}
\caption{$ \dot{R}_{ISCO}$ vs $\alpha$ for $\mu = 10^{-40}$}
\label{fig1}
\end{figure}

\begin{minipage}{.5\textwidth}
  \centering
  \captionsetup{justification=centering}
  \includegraphics[width=\linewidth, height = 6.5cm]{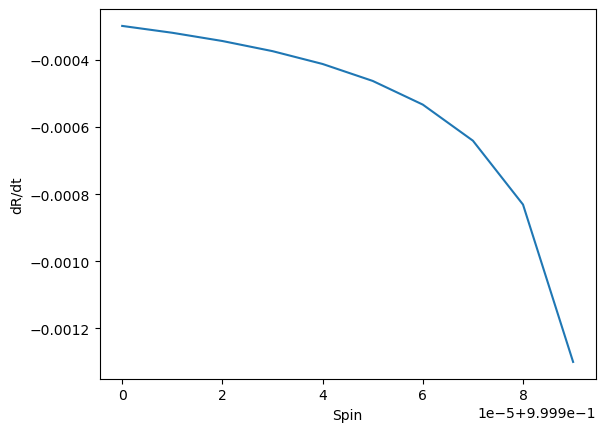}
  \captionof{figure}{$ \dot{R}_{ISCO}$ vs $\alpha$ for $\mu = 10^{-40}$ displaying its transition phase}
  \label{fig2}
\end{minipage}%
\begin{minipage}{.5\textwidth}
 \centering
 \captionsetup{justification = centering}
 \includegraphics[width = \linewidth, height = 6.5cm]{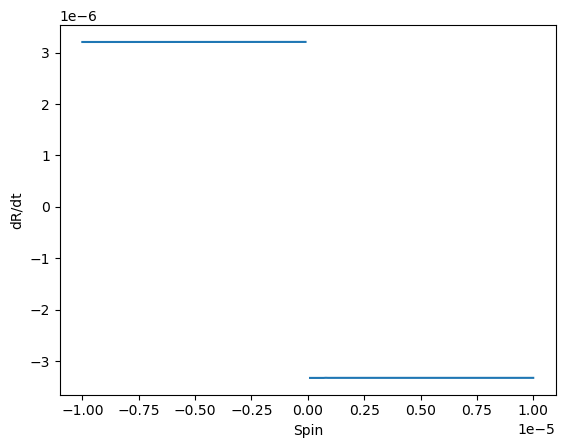} 
 \captionof{figure}{$ \dot{R}_{ISCO}$ vs $\alpha$ for $\mu = 10^{-40}$ displaying discontinuity at $\alpha = 0$}
 \label{fig3}
\end{minipage}
 \begin{figure}[h!]
  \centering
  \includegraphics[width=0.6\linewidth, height  = 6.5cm]{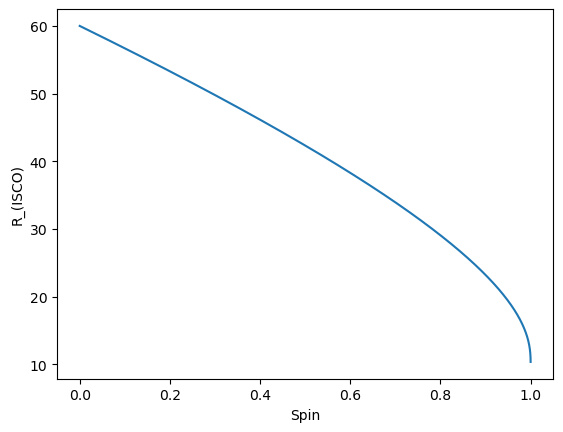}
  \caption{$R_{ISCO}$ vs $\alpha$ for $0<\alpha<1$ and $\mu = 10^{-40}$}
  \label{fig4}
 \end{figure}

\subsection{Large String tension}
For large string tension, we take $\mu = 10^{-29}$ and inserting it into (\ref{string mass loss}) we get $\dot{M}_{str} = 10^{-18} M_{\odot}/yr$. Here we take $\dot{\alpha} = 10^{-10}/yr$ considering the rotational energy extraction by cosmic string slowing down the spin-up process. 

\begin{figure}[!ht]
    \begin{subfigure}{0.5\textwidth}
        \includegraphics[width=1\linewidth, height=6.5cm]{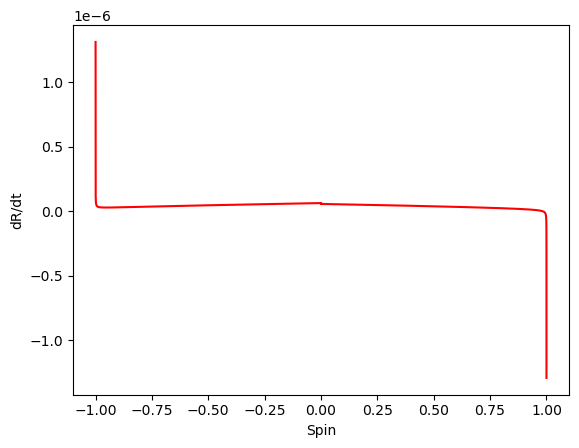} 
        \caption{$-1<\alpha<1$}
        \label{fig5a}
    \end{subfigure}
    \begin{subfigure}{0.5\textwidth}
        \includegraphics[width=1.1\linewidth, height=6.5cm]{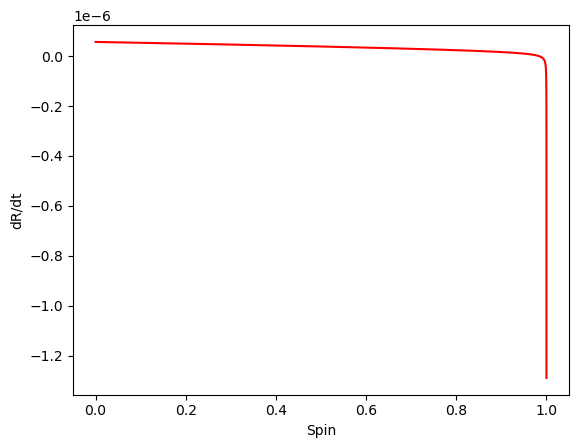}
        \caption{$0<\alpha<1$}
        \label{fig5b}
    \end{subfigure}
\caption{$ \dot{R}_{ISCO}$ vs $\alpha$ for $\mu = 10^{-29}$}
\label{fig5}
\end{figure}

\begin{minipage}{.5\textwidth}
  \centering
  \captionsetup{justification=centering}
  \includegraphics[width=\linewidth, height = 6.5cm]{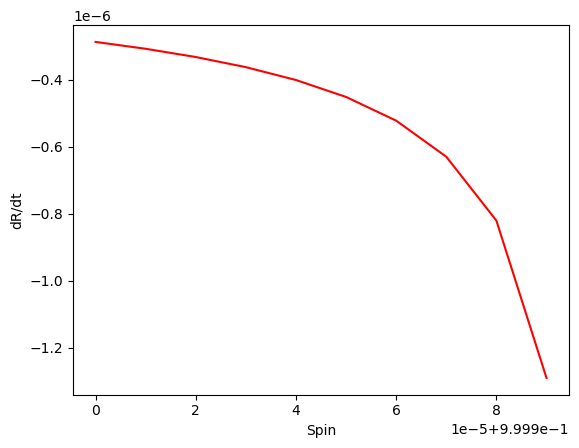}
  \captionof{figure}{$ \dot{R}_{ISCO}$ vs $\alpha$ for $\mu = 10^{-29}$ displaying its transition phase}
  \label{fig6}
\end{minipage}%
\begin{minipage}{.5\textwidth}
 \centering
 \captionsetup{justification = centering}
 \includegraphics[width = \linewidth, height = 6.5cm]{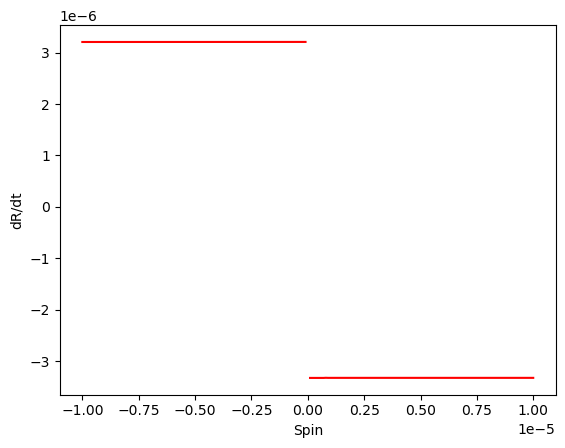} 
 \captionof{figure}{$ \dot{R}_{ISCO}$ vs $\alpha$ for $\mu = 10^{-29}$ displaying discontinuity at $\alpha = 0$}
  \label{fig7}
\end{minipage}
\begin{figure}[h!]  
   \centering
  \includegraphics[width=0.6\linewidth, height = 6.5cm]{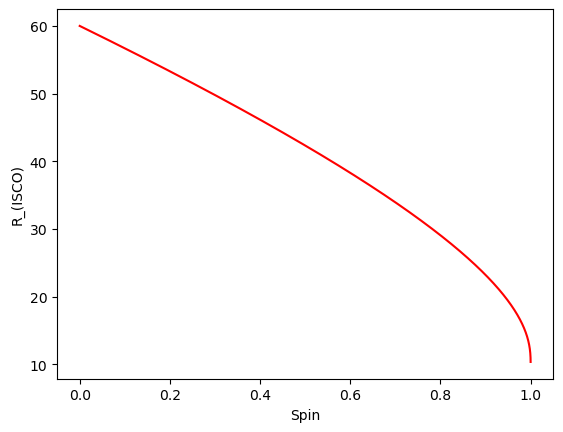}
  \caption{$R_{ISCO}$ vs $\alpha$ for $0<\alpha<1$ and $\mu = 10^{-29}$}
  \label{fig8}
\end{figure}

\section{Results and Discussion}

The first thing to notice while analyzing Fig:\ref{fig1} and Fig:\ref{fig5} is the asymptotic behaviour as $\alpha \to -1$ and $\alpha \to 1$. Interestingly, this behaviour is observed in general, independent of the influence of the cosmic string. To understand the physical interpretation of this, consider a non-rotating black hole that begins to accrete material from the companion star. This process leads to the formation of a disk and an increase in the black hole's spin. The ISCO of a black hole changes when its spin increases due to accretion described by (\ref{general}). However, as the spin reaches its maximum ($\alpha \to 1$), the position of its stable orbit moves closer to the black hole. Once its spin is maximum, the accretion mass gets ejected outwards in the form of winds, analogous to stellar winds caused by a star's rapid rotation. In contrast to the case of prograde motion ($0<\alpha<1$) discussed right above, in the case of retrograde motion($-1<\alpha<0$), the accretion disk spirals into the black hole because the ISCO is now located at infinity at that instant (Fig: \ref{fig1a})(Fig:\ref{fig5a}). It is important to highlight that only the accretion material causing the change in spin is ejected (or falls into the black hole) and the material accreted after maximum spin is attained, does have an ISCO defined by (\ref{bardeen}) which was estimated in \cite{Bardeen72}.  The potential reason could be the force to counter-balance the black hole spin, which abruptly becomes constant at $\alpha = 1$, either through accretion mass falling into the black hole or being ejected outwards. This phenomenon becomes highly observable in the transition phase (Fig:\ref{fig2})(Fig:\ref{fig6}) with rapid change in $ \dot{R}_{ISCO}$ for small changes in $\alpha$. Further, the discontinuity at zero spin (Fig:\ref{fig3})(Fig:\ref{fig7}) plays a role when the black hole is switching between prograde to retrograde motion (or vice-versa). At this transition point, the accretion disk will be ejected outwards in the form of winds. 
\\

Examining the two cases of string tensions depicted in Fig:\ref{fig1b} and Fig:\ref{fig5b}, we notice that although $ \dot{R}_{ISCO}$ for prograde motion in both cases has the same profile, the orders are vastly different. We thus emphasize that, the estimated order $10^{-3}$ for small string tensions (Fig\ref{fig1}) indicates that the location of ISCO is changing at a fast rate, hence providing for an observable effect. However, for the same accretion rates, the presence of a string with large string tension (Fig \ref{fig5}) leads to a lesser change (order $10^{-6}$) indicating that the black hole spin up process is occurring at a slower rate. Such phenomena could indirectly provide for potential evidence of the presence of a cosmic string attached to the black hole. Although this analysis has been conducted over the range $-1<\alpha<1$,  it is crucial to note that in the case of large string tension with $\mu \gtrsim (\dot{M}_{acc})_G$ (Subscript G representing  $\dot{M}_{acc}$ in geometric units where it is dimensionless), the spin of a black hole saturates at $\alpha \sim  (\dot{M}_{acc})_G/\mu$ and never reaches 1 \cite{Xing2021}. This implies that  $\dot{R}_{ISCO}$ never reaches the transition phase (Fig:\ref{fig2}) and therefore does not lose its accretion disk, in contrast to the case with small string tension. 
\\

Specifically, in the case of LMXBs, extremely large cosmic strings, for ex., with $\mu = 10^{-19}$, results in mass loss equivalent to the accretion rate, leading to absence of a spin-up process. If the black hole has a non-zero spin prior to being attached to the cosmic string, strings larger than  $\mu = 10^{-19}$ will lead to the spin-down of a black hole. This spin-down rate can be observed through changes in its ISCO and ultimately resulting in the ejection of accretion mass via winds. 

\section{Conclusions}
In this paper, we have demonstrated that cosmic strings linked to an accreting black hole influence its orbital configuration, by establishing a direct correlation between the ISCO radius and the string tension. Furthermore, it was illustrated that regardless of the presence of cosmic strings, there exists an asymptotic behavior at maximum spin for both prograde and retrograde motions. Building on this observation, the idea has been put forward that the material from the accretion disk is ejected in the form of winds (or spirals inward) during the black hole's spin-up process. Additionally, the discontinuity occurring at zero spin also reflects a similar ejection process (or inward spiral) when the black hole transitions between prograde and retrograde motion (or vice versa).
\\
Cosmic strings with different string tensions were studied in this system emphasizing that its effect on the spin, and consequently the accretion disk structure is significant. We highlight here that the transition phases are the most important phases for observations, due to extreme change in $R_{ISCO}$ for small spin changes. Furthermore the absence of a transition phase also serves as evidence, indicating the existence of a large cosmic string slowing the spin-up process.
\\
We point out here that the accretion disk, consisting of hot plasma and thermal radiations, could influence the structure of the cosmic string. Accreting black holes with jets will also be affected by cosmic strings and conversely, the jets might have an impact on the cosmic string as well. We will address these interactions in our future work.
\\
Current techniques for detecting cosmic strings have yet to find conclusive evidence, and hence we propose this new methodology for detecting them, due to the abundance of XRBs and the significant changes a stellar mass black hole undergoes when attached to a cosmic string. This work is not only an attempt at a novel cosmic string detection technique but also suggests the possibility of mass ejection through winds during the black hole spin-up process in an XRB.
\section{Acknowledgement}
The authors acknowledge Ms. Niyukti Patil for her contribution in Fig \ref{fig}.
\section{References}
\providecommand{\newblock}{}

\end{document}